# Machine Learning Enabled Force Sensing of a Smart Skin for Robotics

*Fan Liu, Guangyu He, Xihang Jiang, and Lifeng Wang\**

Department of Mechanical Engineering, Stony Brook University, Stony Brook, NY 11794, USA

\* Corresponding authors. E-mail address lifeng.wang@stonybrook.edu

**Abstract:**

Artificial skin with the sense of touch can support robots to interact with the surrounding environment efficiently and accomplish complex tasks. Traditional multi-layered artificial skins require complex manufacturing processes which can result in high cost as well as limitations on the material and size of the skin. In this paper, we demonstrate a machine learning based approach to predict positions of point loads using the most direct response as input signal: strain distribution. Starting with the simplest problem, predicting the position of a single point load acting on a flat surface, an ML model is developed, trained, and tested. Accurate predictions are obtained from the ML model, parameters that affect the accuracy are discussed, and validation tests are performed. After that, the ML model is modified to solve multi-objective prediction problems: predicting positions and magnitudes of multiple point loads. In the end, the ML model is upgraded to a 2-step model to predict the position of a point load acting on a deformable surface. The demonstrated approach enables a normal untreated surface to feel a touch no matter what the surface is made of or how large or small the size of the surface is. Therefore, we believe this ML-based load position prediction approach could be a promising tool for applications such as flexible touch screens, smart skin for robots, and micro touch sensors.

## 1. Introduction

The skin is a complex multi-layer structure that can sense pressure, temperature, humidity, and many other environmental stimuli[1]. Combined with sight, hearing and other senses, the comprehensive properties of human sensing help us to efficiently interact with the surrounding environment. Most robots are equipped with high-resolution cameras as eyes and Hi-Fi microphones as ears, however, robot equipped with artificial skin is rarely seen. Without the sense of touch, many tasks, which are easy for human beings, are almost impossible for robots[2]. For example, in grasping objects, the sense of touch can detect important information such as the normal and tangential forces and therefore helps to grasp fragile objects without breaking them. Besides grasping objects, the sense of touch can help get the information of the shape of an object, such as finding a dent on a flat surface. Smart skin with the sense of touch can support robots to accomplish more complex tasks and enhance the potential applications of robots in medical and industrial areas[3]. Because of that, lots of efforts have been made in designing and fabricating smart skin for robots[4].

Recent smart skins are designed with different types of sensing systems including resistive[5], capacitive[6], piezoelectricity[7], and acoustic wave[8]. All these various types of sensing systems have a similar detection process. The touch on the surface creates a signal that is detected and recorded by sensors. Then, the position of the touch is calculated based on the recorded signal[9]. For example, in resistive smart skin, the touch makes the two conductive layers of the skin touch with each other and create an electrical current change which is the signal. The current change is used to calculate the position of the touch. Although these sensing systems achieved great success, each of them has its limitations. The conventional resistive touch system can only detect one touch position. The capacitive touch system can easily achieve a multi-touch function, but it can't detect the magnitude of the touch force precisely. Moreover, most of these techniques need surfaces that are carefully designed and fabricated with multilayers, so they can't be used on normal untreated surfaces.

For a simple flat surface, with a point load acting on it, the most direct response can be obtained is the strain distribution. No matter what the material the surface is made of or how large or small the size of the surface is, point loads acting on different positions of the surface create different strain distributions which, in turn, can be used to calculate the positions of the loads. Different numbers, positions, and magnitudes of the point loads lead to different strain distributions which means the strain distribution contains the information of how many loads acting on the surface, where these loads acting on, and what is the magnitudes. Strain distribution can be easily measured using strain gauges or digital image correlation (DIC) systems. However, the real challenge is using the strain distribution to calculate the location of the point loads.

Machine Learning (ML) was first introduced by Arthur Samuel back in the 1950s when he made the world's first successful self-learning program[10]. Owing to the development of novel machine learning theories and the rapid progress of computer hardware, ML achieved significant success in speech recognition[11], image recognition[12], automated self-driving cars[13], and recommendation engine[14]. And that inspired researchers to apply ML learning to various physical problems. In quantum physics, supervised and unsupervised learning have been used to identify various phases of matter and the transitions between them[15]. Also, by applying artificial neural networks to represent quantum states, researchers have made great progress in solving quantum many-body problems[16]. Moreover, many researchers applied ML in the design and optimization of composites and metamaterials[17]. These studies demonstrate that ML is a powerful tool in various physical fields for its ability to capture the highly complex internal correlations between the different physical variables. Therefore, its highly possible that the internal correlations between strain distribution and point loads can also be captured by a well-designed machine learning model.

In this work, we focus on developing a machine learning based approach to predict the position of point loads on a surface using strain distribution as input. We first discuss the four main steps of developing an ML model and use the model to solve the simplest problem: position prediction of a single point load on a flat surface. Then we perform a validation test and discuss the factors that affect the accuracy of prediction. After that, we modify the model to solve more complex problems: prediction of multiple point loads, prediction of multiple point loads with different load magnitudes. Finally, we create a new 2-step model and solve the biggest challenge: the perdition of point load on a deformable surface.

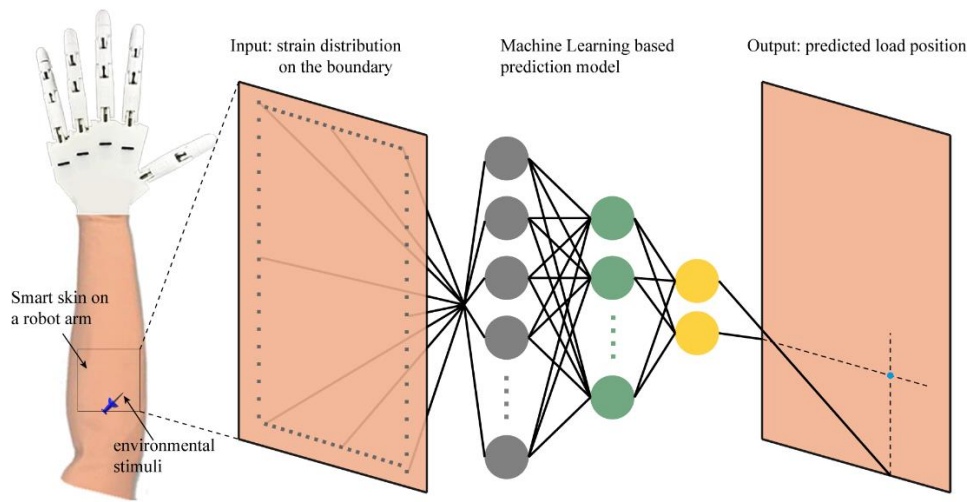

**Figure 1.** Schematic of the machine learning based approach using strain distribution as input data to predict the position of a point load.

## 2. Results and Discussion

### 2.1. Single Point Load Prediction

For a strain-based smart skin, the detected strain distribution is used as an input signal to predict the position of a point load acting on it. Multiple well-developed techniques can be used to obtain the input signal, however, establishing the complex internal correlations between the strain distribution and load position is challenging. Here a supervised machine learning model is an appropriate tool considering the position prediction is a regression problem. The supervised learning model uses a training dataset to teach the model and then produces an inferred function. Therefore, gathering a training dataset is the first step of constructing the ML model.

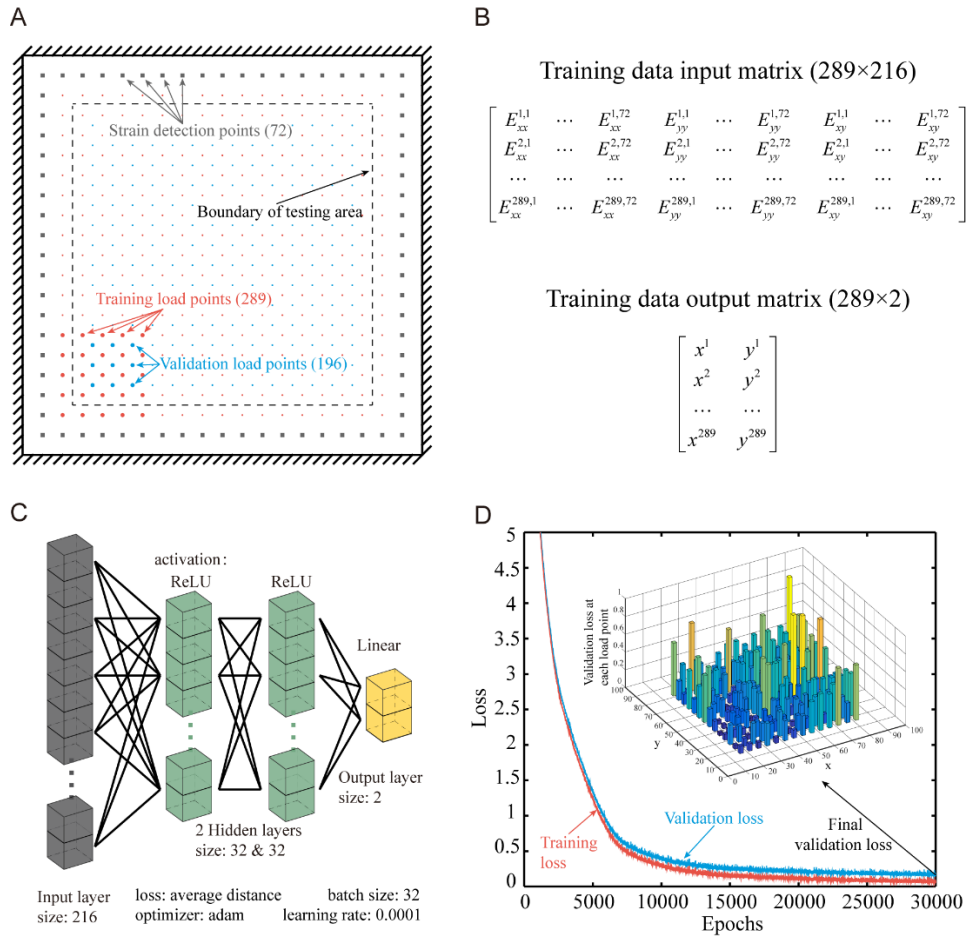

**Figure 2.** A) Finite Element model of point load position prediction problem. B) Input and output matrix of training data obtained from FE simulation results. C) Optimal ML model that gives the most satisfactory results after hyperparameter tuning. D) Training and validation losses during the training process and the final local validation losses of the 196 validation load points.

The training dataset is obtained from finite element simulation results. The FE simulation is performed on a plate (**Figure 2**A) with a normalized size of 100×100 and a normalized thickness of 1. All 4 edges of the plate are clamped, and a point load is acting perpendicular to the plate at one of the red points. The gray blocks on the boundary of the plate are strain detection points, the strain distribution at these points is extracted from the FE simulation results and then used in the ML model. For each FE simulation, an input-output pair is obtained, where the input is the strain components ($E_{xx}$, $E_{yy}$, $E_{xy}$) detected from the 72-strain detection points and the output is the position ($x$, $y$) of the point load. There are a total of 289 red load points for the training dataset on the plate and 289 independent FE simulations are performed to get the training dataset. The training dataset is shown in **Figure 2**B, it has a training data input matrix with a size of 289×216 and a training data output matrix with a size of 289×2. The validation dataset is obtained in the same way and will be used for hyperparameter tuning of the ML model.

The second step of constructing the ML model is developing a first working model. In this step, four key features of the ML model need to be determined and the first one is the type of model. Simple sequential networks, convolutional neural networks, and generative adversarial networks are frequently used in previous studies. In this letter, all three types of models are tried and the simple sequential model is selected for its better performance. Then, the loss function needs to be determined, which is the average distance between true position and predicted position of the point load:

$$Loss = \frac{1}{n}\sum_{i=1}^{n}\sqrt{(x_{true}^i - x_{pred}^i)^2 + (y_{true}^i - y_{pred}^i)^2} \qquad (1)$$

Where *n* is the number of samples in the dataset, *n* is 289 and 196 for the training and validation dataset respectively. The last two features are the optimizer and the activations between layers. Here, linear activation is used in the output layer since it's a regression problem while ReLU activation is used in each hidden layer for its simplicity and high efficiency. As for the optimizer, adam optimizer is used in this ML model. After determining these 4 features, the first working model is constructed with 1 hidden layer. This model is able to predicate the load position, however, it cannot give an accurate prediction because the number and the size of the hidden layers can dramatically affect the performance. Therefore, hyperparameter tuning is essential to find the optimal combination of hyperparameters that minimizes a predefined loss function to give better results.

The third step is tuning the hyperparameters and training the ML model. Here, two hyperparameters are tuned, the number of hidden layers and the size of each hidden layer. The detailed hyperparameters tuning results are shown in the supporting material. After the hyperparameters tuning, the optimal model architecture (**Figure 2**C) is obtained, which has 2 hidden layers and each hidden layer has 32 neurons. This ML model is trained for 30000 epochs, and the training loss and the validation loss are calculated at each epoch and shown in **Figure 2D**. The training loss and validation loss both decrease with the increase in the number of epochs. Also, there is no overfitting in the training process. After 30000 epochs, the final loss of the training dataset and validation dataset are lower than 0.3, suggesting the overall prediction is accurate considering the size of the plate is 100×100. Note that the training/validation loss is the average distance between the real and predicted positions of all points in the training/validation dataset. Besides the overall loss, the local loss at each point also needs to be validated in case of the possible extreme high loss at some local points. The local validation losses of the trained model are calculated (**Figure 2**D). The maximum and minimum values are 0.78 and 0.01 respectively, suggesting that the trained model can make accurate predictions with no outliers.

After the three steps discussed above, the well-trained ML model is tested on a randomly generated test dataset to evaluate the performance. Here 100 points on the plate are randomly generated and 100 independent FE simulations with point load acting on these points are performed. Then a test dataset is created by following the same process used for creating the training and validation dataset. The input matrix of the test dataset has the size of 100×216 contains the strain distribution information and the output matrix has the size of 100×2 which are the true positions of the 100 points. By applying the trained ML model on the input matrix, predicted positions of the 100 points are obtained. The true and the predicted positions are plotted in **Figure 3**A which shows straightforwardly the high accuracy of the ML model. Furthermore, a larger test with 3000 points is generated. All 3000 local losses for those points are calculated, and their histogram distribution is shown in **Figure 3**B. The majority (more than 80%) of the points in the test dataset show loss values lower than 0.5 and none of the points has a loss value larger than 1. The results clearly show that the trained ML model has a good performance on the randomly generated test dataset. At this point, the entire process of making an accurate prediction of the position of a point load on a plate based on the information of strain distribution is completed. The accurate prediction results proved that theoretically at least, position prediction of a point load on a flat surface is achievable using a properly designed ML model with strain distribution as the input signal.

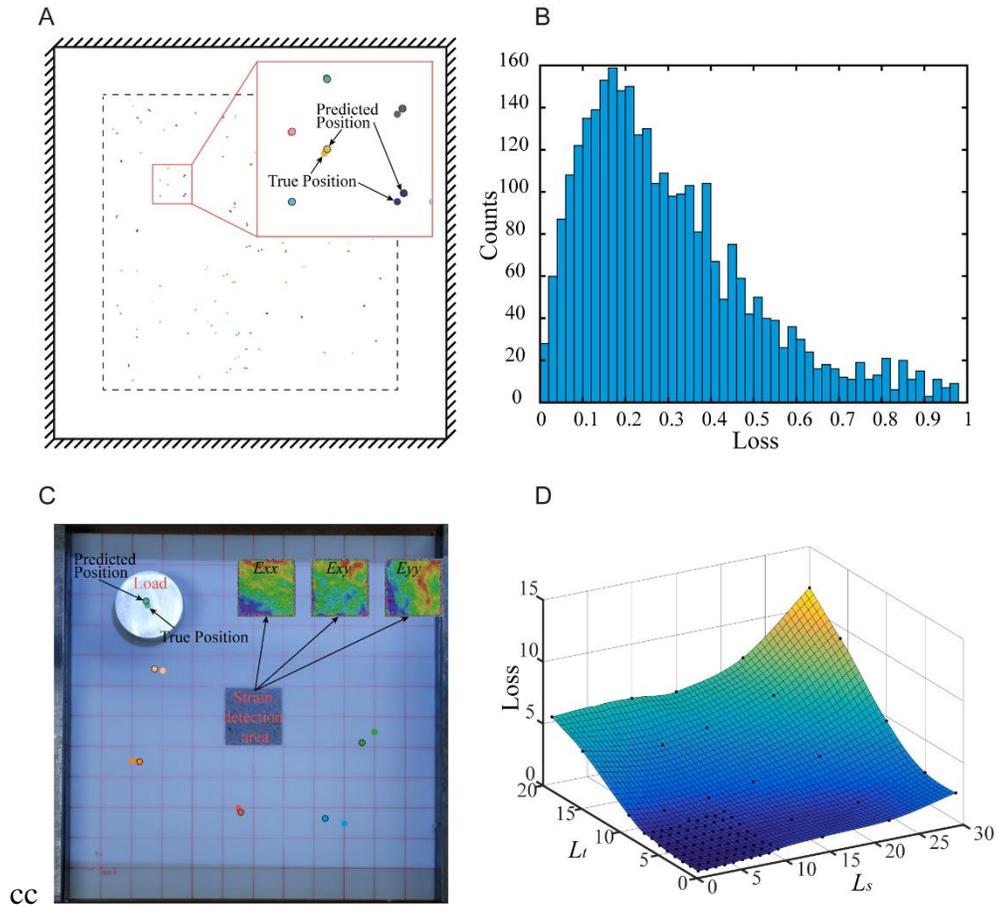

**Figure 3.** A) The true and predicted positions of 100 randomly generated point loads in the test dataset. B) Histogram of the test losses of 3000 randomly generated point loads in the test dataset. C) Results from validation test using strain distribution at the center of the plate. D) The effect of $L_s$ and $L_t$ on the final test loss.

## 2.2. Experimental results and accuracy improvement discussion

To verify the results, a validation test is performed, and the detailed setup and the procedure of the test are discussed in the supporting material. The test results are shown in **Figure 3**C, where the true positions and predicted positions are plotted on the plate. Clearly, the ML model still performs well in the validation test with slightly lower accuracy. The lower accuracy can be attributed to a smaller number of training points and the error from the strain measurement. Knowing how these factors affect the accuracy and then improving the accuracy is critical.

No matter what method is used to measure the strain distribution, DIC, or strain gauge, the measurement error is unavoidable. The measurement error can obviously lead to the increase of the loss. However, the effect can vary for loads with different magnitudes. The relationship between test loss and the measurement error is calculated respectively for loads with different magnitude. The results are shown in **Figure** S4. When measurement error equals 0, the losses converge to a constant value of 0.38. As the measurement error increases, the losses for different loads all increase but at different increase rates. The loss of a large load increases much slower than the loss of a small load. Therefore, for a given measurement error, the location prediction of a large load is more accurate than that of a small load.

The number of strain detection points and training load points can also affect the accuracy. The interval between strain detection points, $L_s$, and the interval between training load points, $L_t$, are used to represent the number of strain detection points and training load points respectively. In the previous model, with $L_s$ and $L_t$ both equal to 5, there are 72 strain detection 289 training load points accordingly. To study the effect of $L_s$ and $L_t$ on the accuracy, a new model with $L_s$ and $L_t$ both equal to 1 is created. The size of the new input matrix is 6561×360 which is much larger than the input matrix of the previous model. After training with the new dataset, the final loss of the new ML model is calculated. The new final loss is 0.06 which is much smaller than the final loss of the previous model. The comparison suggests that increasing the number of strain detection points and training load points can help to improve the accuracy of the model. To get a better understanding of the effect of $L_s$ and $L_t$, more ML models with different $L_s$ and $L_t$ are created and trained. The final losses obtained from these different models are shown in **Figure** 3D. The loss decreases as the $L_t$ decreases and the $L_s$ decreases. This is because a smaller $L_s$ leads to more strain detection points and therefore an input matrix with more columns that contains more information on strain distribution. A smaller $L_t$ leads to more training load points and therefore an input matrix with more rows that contains more samples used for training. Therefore, a larger input matrix contains more strain distribution information and more samples can certainly make more accurate predictions.

### 2.3. Multiple Point Loads Prediction

In the previous section, the well-trained ML model can make an accurate prediction for the position of a single point load. This model can also be used to predict the positions of multiple point loads, but some modifications need to be made. In the previous model, there is one local loss for each sample which is the distance between the true position and predicted position of a point load. Here, with multiple point loads in each sample, there are multiple local losses accordingly: $L_1, L_2, ..., L_m$. Therefore, the overall loss for the model training is:

$$Loss = \frac{1}{n}\frac{1}{m}\sum_{j=1}^{m}\sum_{i=1}^{n}\sqrt{(x_{true}^{i,j} - x_{pred}^{i,j})^2 + (y_{true}^{i,j} - y_{pred}^{i,j})^2} \qquad (2)$$

where $n$ is the total number of samples in the dataset, $m$ is the total number of point loads. Other than the loss function, the training/validation datasets need to be changed as well. In the previous model, there are two columns in the output matrix representing the position $(x,y)$ of one point load. For multiple point loads, the number of the columns in the output matrix is $2m$ representing the positions $(x_1, y_1, x_2, y_2, ..., x_m, y_m)$ of all point loads. The input matrix used for multi-point prediction is the same as the one used in 1-point prediction because the strain distribution information is obtained from the same 72 strain detection points no matter how many point loads acting on the plate. Lastly, for multiple point loads, there are much more samples in the training dataset. With these modifications, we trained 2-point models (results shown in Figure S5) and 3-point models with different model structures. The validation losses during the training process of the 3-point models is shown in **Figure** 4A. In the previous section, after hyperparameter optimization, the model structure of 2×32 (2 layers and 32 neurons in each layer) is selected and achieved high accuracy in the 1-point model. Here, for the 3-point model, the same model structure (2×32) is used but the final loss is very high as shown in **Figure** 4A. It is because the 3-point model is much complex than the 1-point model, so a simple 2×32 structure is too shallow to capture all intrinsic correlations in the 3-point model. By adding more layers and more neurons in each layer, more accurate ML models are created and trained. The results are shown in **Figure** 4A, with a model structure of 8×256, the final loss of that 3-point model reaches 0.43, suggesting that the multi-point prediction can also be highly accurate.

After modification and training of the 3-point model, a test dataset with 100 test samples is randomly generated. The well-trained 3-point model is used to calculate the predicted positions of the 3-point loads in each test sample. The losses between the predicted positions and true positions of all the 100 samples are shown in **Figure** 4B. All loss values are lower than 0.8 which indicates that the 3-point model can make accurate predictions on the test dataset. The loss values of different samples are different for sure, and for each sample, the loss values of different points are also different. In the first sample, the loss of the second point is much smaller compared to the loss of the first and third points. However, in the second sample, the loss of the second point is larger than the loss of the first and third points. To check if the model tends to make a more accurate prediction for 1 point than the other 2, we sum up all losses of the first point in the 100 samples. The summation of the losses of second points and third points are also calculated and the comparison of them is shown in **Figure** 4B. The summations of losses of the three points are almost the same which implies the prediction made by the ML model for the three points have the same accuracy in general.

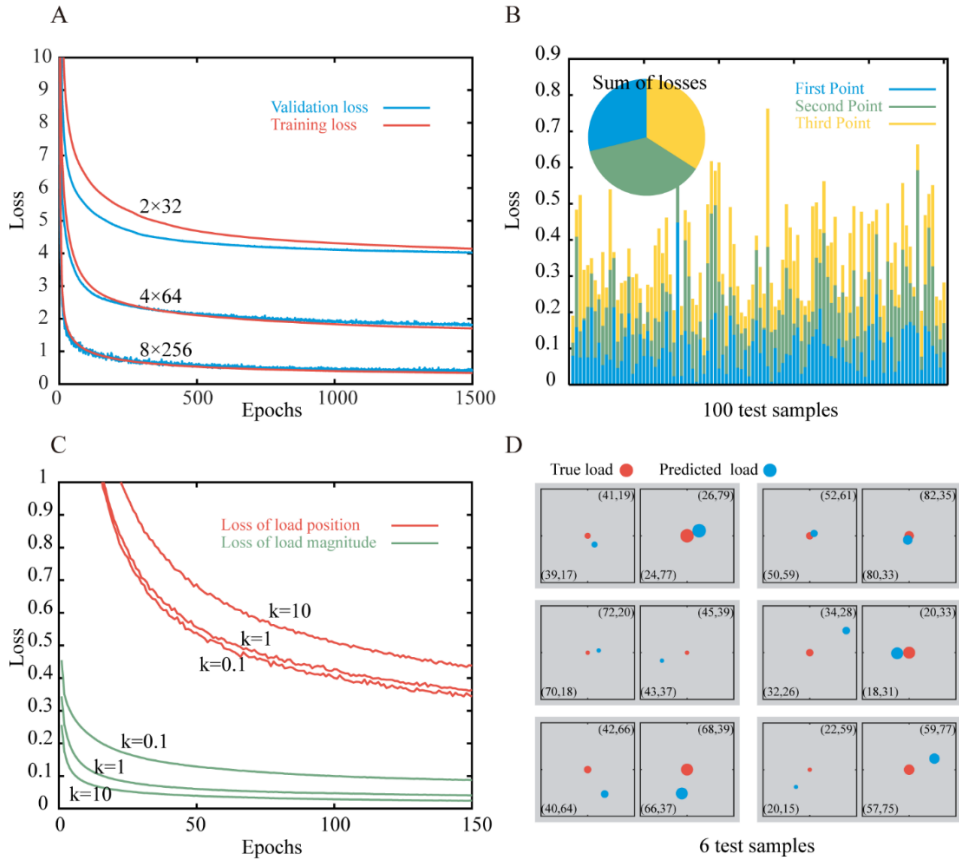

**Figure 4.** A) Training and validation losses during the training process of the 3-point model with structures of 2×32, 4×64, and 8×256. B) The test losses of 100 randomly generated 3-point samples. C) Validation losses during the training process of the 2-point different magnitude model with scale factor $k = 0.1$, 1, and 10. D) The true and predicted positions of 6 randomly generated 2-point different magnitude load samples in the test dataset.

The above results show that the position of one point and multiple points can both be predicted accurately by the ML model. However, in the multiple-point prediction, we simply assume that the magnitudes of the multiple point loads are the same. But in practice, an unavoidable challenge is that the different point loads may have different magnitudes. Therefore, a modified ML model that can predict the positions, as well as the magnitudes of multiple point loads, is needed. Here, we focus on a model for 2-point loads with different magnitudes. The first step of developing this new model is to create new training and validation datasets. Here, the output matrix has 6 columns ($x_1$, $y_1$, $x_2$, $y_2$, $m_1$, $m_2$) which represent the position and magnitude of the two-point loads. There are 289 possible positions and 10 possible magnitudes for each point load which covers the load range from $\Delta_0$ to 10 $\Delta_0$. Therefore, a total of 86400 samples are in the training dataset. The second step is to modify the loss function which is the objective function we aim to optimize during the training process. In the previous section, the loss functions of the single-point model and multiple point model are intrinsically the same: the distance between the true position and the predicted position of the point load. However, for the new model, the loss function has two parts, the loss of position and the loss of magnitude:

$$Loss = \frac{1}{n}\frac{1}{m}\sum_{j=1}^{m}\sum_{i=1}^{n}\left(\sqrt{(x_{true}^{i,j} - x_{pred}^{i,j})^2 + (y_{true}^{i,j} - y_{pred}^{i,j})^2} + \left|\frac{m_{true} - m_{pred}}{m_{true}}\right| \cdot k\right) \quad (3)$$

Where the first part is the loss of load positions which is the average distance between the true and predicted positions. The second part is the loss of magnitude which is the difference between true and predicted load magnitude. $k$ is a scale factor. The structure of the model is determined after hyperparameter tuning, which has 8 layers and 256 neurons in each layer. The training results for models with different $k$ are shown in **Figure** 4C. When $k = 1$, the total loss is simply the summation of position loss and magnitude loss. After 150 epochs of training, the average distance is less than 0.5 while the average magnitude difference is less than 0.4 $\Delta_0$. By changing $k$ from 1 to 0.1, the proportion of magnitude loss in the total loss decreases. Therefore, in the training process, the model tends to give a more accurate position prediction and a less accurate magnitude prediction. On the contrary, the model with $k = 10$ has a more accurate magnitude prediction and a less accurate position prediction. The scale factor $k$ can be used to balance the trade-offs between the accuracy of position and the accuracy of load magnitude.

The test dataset is created in the end which has 100 samples. For each sample, there are 2-point loads with randomly generated position and load magnitude. The input matrix, the strain distribution information, is then used by the well-trained model (with $k = 1$) which makes a prediction of the position and magnitude of the 2-point loads in each sample. Here the first 6 samples are shown in **Figure** 4D. Each small block has a size of 2×2 while the coordinates of the diagonal points are marked on the figure. Red and blue dots are used to represent the true and predicted point loads respectively. The size of the dot is proportional to the load magnitude. The results show that the distances between true and predicted points are very small for most points. The first point in the 4th sample has the most inaccurate position prediction but the distance is still less than 1. As for the accuracy of magnitude prediction, we can hardly tell the difference by comparing the size of red and blue dots.

**2.3. Point Load Prediction in a Deformable Surface**

The results in section 2.1 demonstrate that our ML model can predict the position of a single point load. But it has been achieved decades before using many other methods. Then, in section 2.2, modified models successfully achieved the multiple points prediction and load magnitude prediction. However, these are nothing new in the last ten years. A more complex and interesting topic is making load position prediction on a deformable surface which has been rarely studied. **Figure** 4A shows two basic types of deformation of a surface: bending and torsion. Two parameters are used to define the bending and torsion deformation: $\Delta l$ and $\Delta \theta$. Here, we assume the deformation of the surface is simply the combination of bending and torsion. Moreover, we also assume the surface is smooth, so the load direction is perpendicular to the surface.

In the previous study, the point load is the only factor that contributes to the strain distribution. However, in this problem, strain distribution is induced by the point load and as well as surface deformation. The surface is flat at the beginning, then the surface deformation creates a strain distribution which is named Strain_B&T. After the deformation, a point load is acting on the deformed surface, the strain distribution now is called Strain_L. Strain_L is created by the deformation and point load together, therefore, it contains the information of surface deformation and load positions both. Because of that, theoretically, an ML model uses Strain_L as input data only is able to predicate the load position. Here, an ML model using Strain_L as input data is created. The output matrix of the model has four columns, ($x$, $y$, $\Delta l$, $\Delta \theta$), the first two define the position of the load and the second two define the deformed shape of the surface. New FE simulations on the deformed surface are carried out to generate the training and validation dataset. Here, the loss function of the model is:

$$Loss = \frac{1}{n}\sum_{i=1}^{n}\left(\sqrt{(x_{true}^i - x_{pred}^i)^2 + (y_{true}^i - y_{pred}^i)^2} + \left(\left|\Delta l_{true} - \Delta l_{pred}\right| + \left|\Delta \theta_{true} - \Delta \theta_{pred}\right|\right)\cdot k\right) \quad (4)$$

Where the first part is the loss of load positions and the second part is the loss of surface deformation and $k$ is a scale factor. After the hyperparameter tuning, the structure of the model is set as 64×4. The training results are shown in **Figure** 5C. The final validation loss is around 1.0.

In the previous section, the trade-off between position and load magnitude is unavoidable because the position and magnitude can both affect the strain distribution. Also, their effects on the strain distribution cannot be easily separated. But here, the trade-off problem between surface deformation and load position can be solved by introducing a 2-step model.

In the 2-step model, two sub-models are used sequentially as shown in **Figure** 5B. In the first step, the Strain_B&T matrix is used as the input data in sub-model 1. Then the bending and torsion deformation, ($\Delta l$, $\Delta \theta$), is predicted by the sub-model 1 as output data. Lastly, sub-model 2 uses the predicted surface deformation together with strain_L as input data to predict the load position. The optimized structure of sub-model 1 and sub-model 2 are 64×4 and 128×4 respectively. The training and validation losses during the training process of the 2-step model are shown in **Figure** 5C and the 2-step model achieves higher accuracy than the 1-step model.

The last step is to verify these two models using randomly generated test datasets. 100 samples for the test dataset are randomly created using FE simulation. For the 1-step model and 2-step model, the final losses of the test dataset are 0.48 and 1.58 respectively. The 2-step model has much higher accuracy. In the FE simulation used to generated the training dataset, the magnitude of the point load is set to $\Delta_0$. Therefore, if the load magnitude in the test dataset is higher or lower than $\Delta_0$, the results will be affected. More groups of test datasets are generated using different load magnitudes in the FE simulation, where the range of the magnitude is from $0.5\Delta_0$ to $1.5\Delta_0$. The losses of these datasets are calculated and plotted in **Figure** 5D. The results of both two models are affected because of the change of the load magnitude, however, the 2-step model has much stable and accurate results compared to the 1-step model.

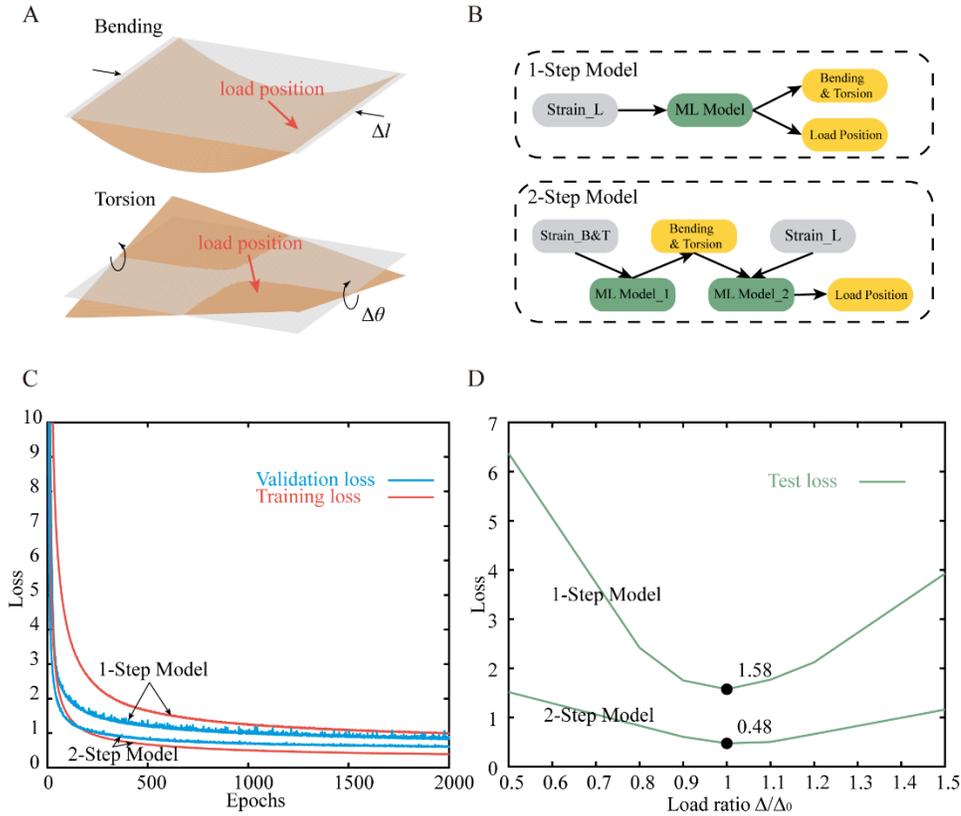

**Figure 5.** A) Two basic types of deformation of a surface: bending and torsion. B) The workflow of the 1-Step model and 2-Step model. C) Training and validation losses during the training process of the 1-Step model and 2-Step model. D) Test losses of 1-Step model and 2-Step model with different load ratios.

## 3. Conclusion

In this paper, we have developed an approach to predict the positions of point loads acting on a surface using an ML-based prediction model. We started with the simplest problem: position prediction of a single point load on a flat surface. The complete procedure of ML-based prediction is discussed in detail which includes four main steps: 1) Generate the training dataset from finite element analysis simulation or experiment; 2) Choose the proper type of networks, loss function, optimizer, and the activations; 3) Tune the hyperparameters and train the machine learning model; 4) Generate test dataset and validate the results. Other than performing analysis based on simulation results, a validation test is performed, and the results confirmed that the ML-based model can provide an accurate prediction in the test. Also, we discussed the factors that affect the accuracy of the perdition results and proved that the model can achieve higher accuracy with more training samples and more strain detection points. Moreover, we modified and upgraded the model and used it on more complex problems. The modified ML models performed well in prediction positions and magnitudes of multiple point loads. For the problem of predicting the load position on a deformable surface, a 2-step model is developed and accurate results are obtained.

The accurate prediction in this study can be attributed to the ability of the ML model that can capture the highly complex inherent relationship between the load position and the most direct response of the surface: the strain distribution. Considering the only information needed is the strain distribution, this prediction approach has 2 important advantages. First, the surface itself does not need any special treatment, so this approach can be used on all types of surfaces made of different materials. Second, the size of the surface does not affect the results, so this approach can be used on surfaces with different length scales: from meter down to nanometer scale. Because of that, we believe this ML-based load position prediction approach could be a promising tool for applications such as flexible touch screens, smart skin for robots, and micro touch sensors.

## 4. Experimental Section

*Numerical Simulations:* The finite element simulations were carried out using the commercial FE package Abaqus to generate the datasets for machine learning models. Models were built using four-node shell elements (S4R) with 5 integration points in the thickness direction, and the strain distributions were obtained from section point 1 (top surface). A linear elastic material model was used in the simulation. Also, the finite element model was verified by a mesh sensitivity test and there were 1000000 elements in each model. The Abaqus/Standard solver was employed for all simulations and geometric nonlinearities were considered in the simulations in section 2.3 because of the large deformation of the surface.

*Machine Learning Model:* The ML models were designed and implemented using an open-source neural-network library TensorFlow. Fully connected neural networks (FCNNs) were adopted in this work, specifically, in section 2.3, the 2-step model has 2 FCNNs running sequentially. The number of layers and the neurons in each layer was determined by hyperparameter tuning to achieve high accuracy within a reasonable compute time. The hyperparameter tunings were performed using an open-source library KerasTuner with the search algorithm of Random Search. The activation function is a rectified linear unit (ReLU) with different self-defined loss functions for corresponding models. Adam optimizer was used to update the weights to minimize the loss function based on analytical gradients calculated in backpropagation. The ML models are trained using the Nvidia Tesla K80 GPUs and the compute time for the most complex model, 8 layers and 256 neurons in each layer for 3-point position prediction, is about 70 hours.

*Validation Test:* The validation test was performed on a PVC plate with the size of 10 inches × 10 inches and 1mm in thickness. Two edges of the plate are clamped. The ML model is created and trained following the same procedure. Digital Image Correlation (DIC) measurement system was used for obtaining the training dataset based on the deformed speckle pattern in the middle of the plate. The ML model for the validation test was trained and saved. After that, with a point load applied at any position on the plate, the predicted position was calculated using the well-trained model and shown on the plate using a projector installed above the plate. The detailed process can be found in the supporting video.

**Supporting Information**

Supporting Information is available from the Wiley Online Library or from the author.

**Acknowledgments**

The authors would like to thank Stony Brook Research Computing and Cyberinfrastructure, and the Institute for Advanced Computational Science at Stony Brook University for access to the high-performance SeaWulf computing system, which was made possible by a $1.4M National Science Foundation grant (#1531492).

**References**

[1]     G. Y. Bae, J. T. Han, G. Lee, S. Lee, S. W. Kim, S. Park, J. Kwon, S. Jung, K. Cho, *Adv. Mater.* **2018**, 30, 1803388.

[2]     a) M. Li, K. Hang, D. Kragic, A. Billard, *Robot. Auton. Syst.* **2016**, 75, 352; b) J. Mahler, M. Matl, V. Satish, M. Danielczuk, B. DeRose, S. McKinley, K. Goldberg, *Sci. Robot.* **2019**, 4.

[3]     C. Bartolozzi, L. Natale, F. Nori, G. Metta, *Nat. Mater.* **2016**, 15, 921.

[4]     X. Wang, L. Dong, H. Zhang, R. Yu, C. Pan, Z. L. Wang, *Adv. Sci.* **2015**, 2, 1500169.

[5]     a) R. Aguilar, G. Meijer, presented at SENSORS, 2002 IEEE **2002**; b) D. S. Hecht, D. Thomas, L. Hu, C. Ladous, T. Lam, Y. Park, G. Irvin, P. Drzaic, *J. Soc. Inf. Disp .* **2009**, 17, 941.

[6]     a) P. T. Krein, R. D. Meadows, *IEEE Trans. Ind. Appl.* **1990**, 26, 529; b) J.-Y. Ruan, P. C.-P. Chao, W.-D. Chen, presented at SENSORS, 2010 IEEE **2010**.

[7]     a) K. I. Park, J. H. Son, G. T. Hwang, C. K. Jeong, J. Ryu, M. Koo, I. Choi, S. H. Lee, M. Byun, Z. L. Wang, *Adv. Mater.* **2014**, 26, 2514; b) Z. L. Wang, J. Song, *Science* **2006**, 312, 242.

[8]     R. Adler, P. J. Desmares, *IEEE T. Ultrason. Ferr.* **1987**, 34, 195.

[9]     M. R. Bhalla, A. V. Bhalla, *Int. J. Comput. Appl.* **2010**, 6, 12.

[10]    A. L. Samuel, *IBM J. Res. Dev.* **1959**, 3, 210.

[11]    L. Deng, J. Li, J.-T. Huang, K. Yao, D. Yu, F. Seide, M. Seltzer, G. Zweig, X. He, J. Williams, presented at 2013 IEEE International Conference on Acoustics, Speech and Signal Processing **2013**.

[12]    G. Koch, R. Zemel, R. Salakhutdinov, presented at ICML deep learning workshop **2015**.

[13]    S. Grigorescu, B. Trasnea, T. Cocias, G. Macesanu, *J. Field Robot.* **2020**, 37, 362.

[14]    S. Zhang, L. Yao, A. Sun, Y. Tay, *ACM Comput. Surv.* **2019**, 52, 1.

[15]    B. S. Rem, N. Käming, M. Tarnowski, L. Asteria, N. Fläschner, C. Becker, K. Sengstock, C. Weitenberg, *Nat. Phys.* **2019**, 15, 917.

[16]    G. Carleo, M. Troyer, *Science* **2017**, 355, 602.

[17]    a) C. T. Chen, G. X. Gu, *Adv. Sci.* **2020**, 7, 1902607; b) G. X. Gu, C.-T. Chen, D. J. Richmond, M. J. Buehler, *Mater. Horiz.* **2018**, 5, 939; c) F. Liu, X. Jiang, X. Wang, L. Wang, *Extreme Mech. Lett.* **2020**, 41, 101002.

Supporting Information

**Machine learning based smart skin for robotics:**

**enable a normal surface to feel a touch**

*Fan Liu, Guangyu He, Xihang Jiang, and Lifeng Wang\**

**Table S1.** Hyperparameters tuning results of the top 50 models.

| rank | loss | layer-1 | layer-2 | layer-3 | layer-4 |
|------|------|---------|---------|---------|---------|
| 1 | 0.24 | 48 | 8 | 128 | 128 |
| 2 | 0.24 | 4 | 64 | 128 | 64 |
| 3 | 0.25 | 4 | 48 | 128 | 32 |
| 4 | 0.26 | 32 | 64 | 32 | 32 |
| 5 | 0.26 | 128 | | | |
| 6 | 0.27 | 32 | 32 | | |
| 7 | 0.27 | 32 | 128 | 8 | 128 |

| rank | loss | layer-1 | layer-2 | layer-3 | layer-4 |
|---|---|---|---|---|---|
| 8 | 0.27 | 128 | 24 | | |
| 9 | 0.28 | 64 | 4 | 24 | 8 |
| 10 | 0.28 | 48 | 48 | | |
| 11 | 0.28 | 128 | 16 | | |
| 12 | 0.28 | 128 | 24 | 32 | 32 |
| 13 | 0.28 | 48 | 128 | 4 | 128 |
| 14 | 0.29 | 16 | 32 | 8 | 48 |
| 15 | 0.29 | 8 | 48 | 24 | 48 |
| 16 | 0.29 | 32 | 128 | | |
| 17 | 0.29 | 128 | 16 | 8 | 64 |
| 18 | 0.29 | 24 | 128 | 4 | 64 |
| 19 | 0.29 | 24 | 16 | 32 | 24 |
| 20 | 0.29 | 48 | 128 | 24 | 16 |
| 21 | 0.30 | 128 | 48 | 4 | 128 |
| 22 | 0.30 | 48 | 32 | 4 | 48 |
| 23 | 0.30 | 64 | 16 | 64 | 64 |
| 24 | 0.30 | 64 | 4 | 128 | 8 |
| 25 | 0.30 | 16 | 24 | 48 | 48 |
| rank | loss | layer-1 | layer-2 | layer-3 | layer-4 |
| 26 | 0.30 | 24 | 48 | | |
| 27 | 0.30 | 32 | 64 | 24 | 128 |
| 28 | 0.31 | 64 | 16 | 32 | 128 |
| 29 | 0.31 | 64 | 32 | | |
| 30 | 0.31 | 48 | 64 | 64 | 32 |
| 31 | 0.36 | 8 | 24 | 24 | 128 |
| 32 | 0.37 | 24 | 8 | 24 | 48 |
| 33 | 0.37 | 4 | 64 | 8 | 24 |
| 34 | 0.38 | 32 | 8 | 8 | 24 |
| 35 | 0.39 | 32 | 8 | 4 | 4 |
| 36 | 0.40 | 64 | | | |
| 37 | 0.40 | 32 | 8 | | |
| 38 | 0.41 | 32 | 4 | 32 | 4 |
| 39 | 0.42 | 8 | 48 | | |
| 40 | 0.44 | 4 | 24 | 24 | 48 |
| 41 | 0.44 | 4 | 4 | 128 | 32 |
| 42 | 0.53 | 24 | | | |
| 43 | 0.53 | 4 | 24 | 32 | 32 |
| 44 | 0.54 | 8 | 8 | | |
| 45 | 0.58 | 8 | 16 | | |
| 46 | 0.72 | 8 | 8 | | |
| 47 | 0.84 | 16 | | | |
| 48 | 0.86 | 4 | 16 | | |
| 49 | 1.12 | 8 | | | |
| 50 | 1.38 | 4 | | | |

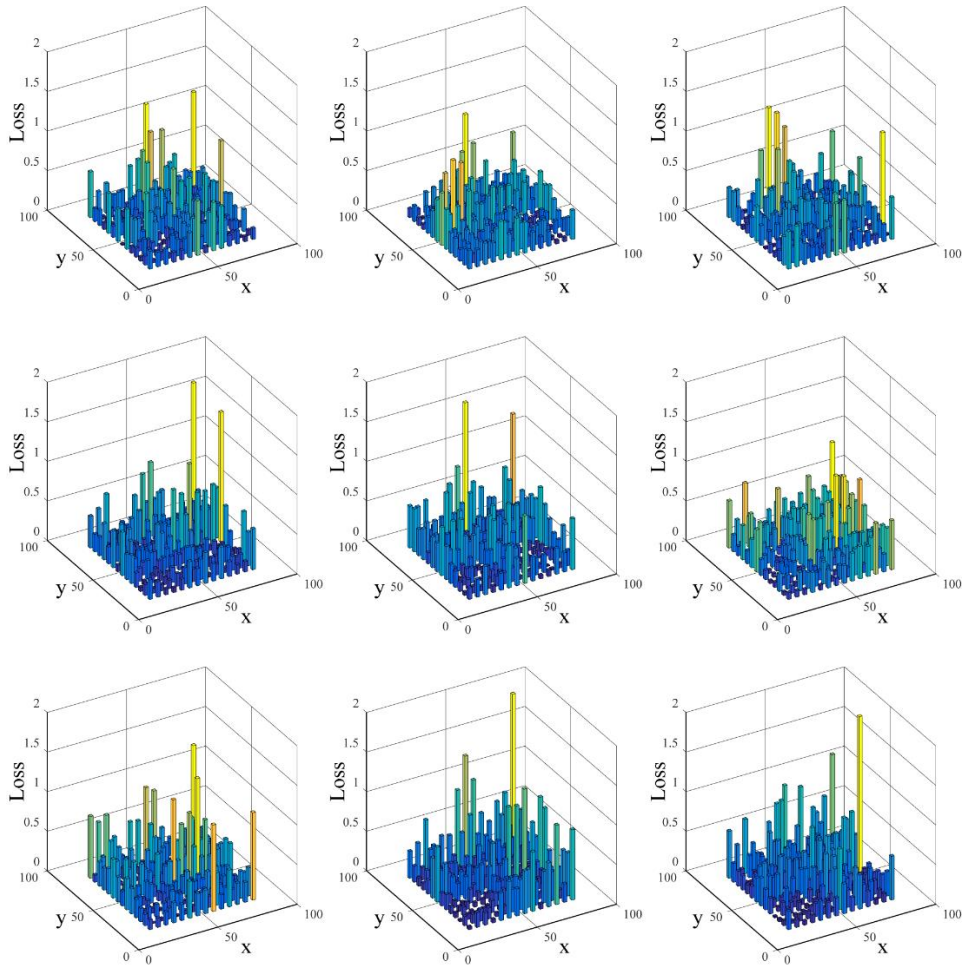

**Figure S1.** The local validation losses of the top 9 models in hyperparameters tuning. For models with different structures, the local validation losses distributions are different. There are several high-loss points in each model, and the prediction accuracy near these points is relatively lower. In the ML model, the overall loss is used to evaluate the accuracy of the model, but the local losses also need to be taken into consideration in case there is a highly inaccurate point. The 6$^{th}$ model is chosen because the highest local loss in that model is still less than 1, which is much better than the top 5 models.

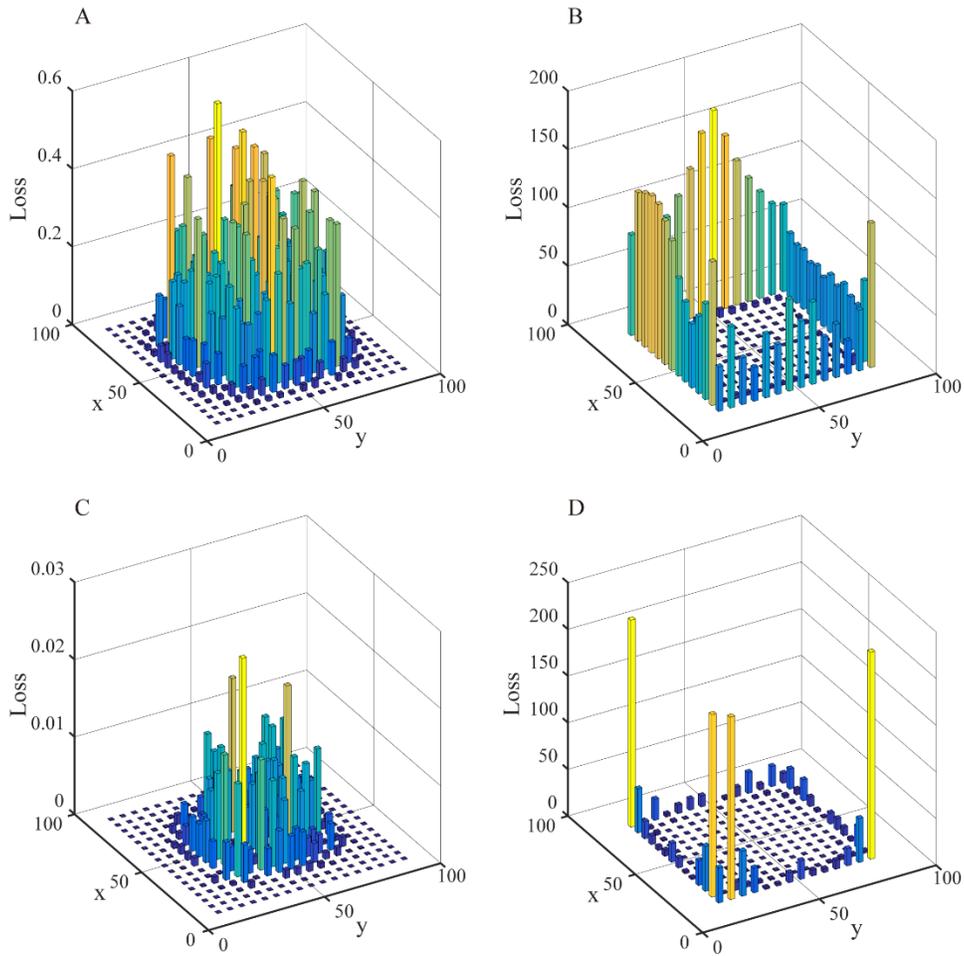

**Figure S2.** A-B) Training and validation losses using simple linear regression. C-D) Training and validation losses using quadratic regression. For the single-point position prediction problem, the ML model is used to establish the internal correlations between the strain distribution and load position. The question seems quite simple that traditional regression models may also perform well. However, after the calculation, it is clear that the regression model cannot provide accurate prediction especially on the boundary of the plate.

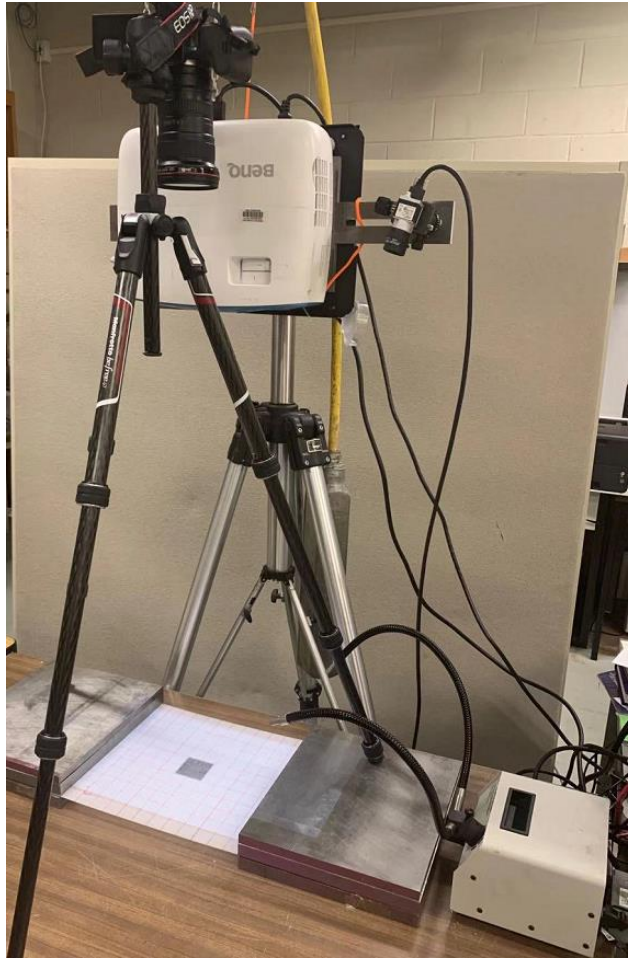

**Figure S3.** Experimental setup of the validation test. The test is performed on a PVC plate with the size of 10 inches × 10 inches. Two edges of the plate are clamped. The ML model is created and trained following the same procedure. The only difference is that the training dataset is obtained using the DIC technique based on the deformed speckle pattern in the middle of the plate. After obtaining a well-trained model, the point load can be applied at any position on the plate. The predicted position is then calculated and shown on the plate using a projector. The detailed process can be found in the supporting video.

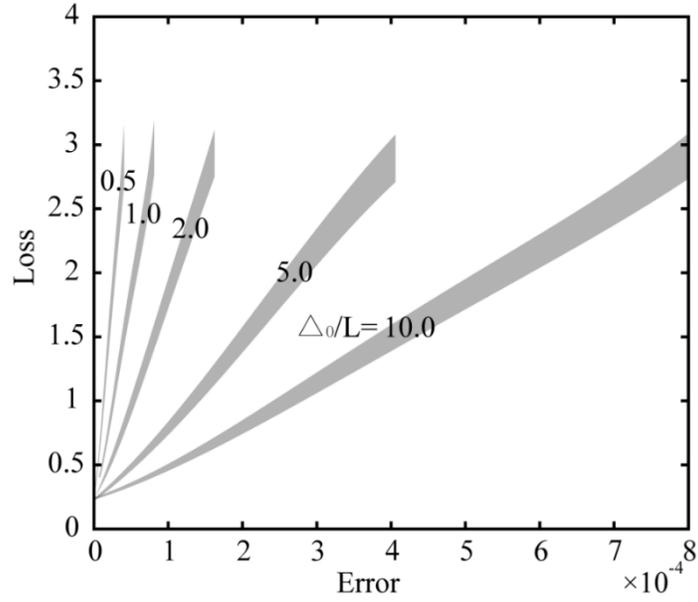

**Figure S4.** Test losses versus measurement error for models with different load magnitudes.

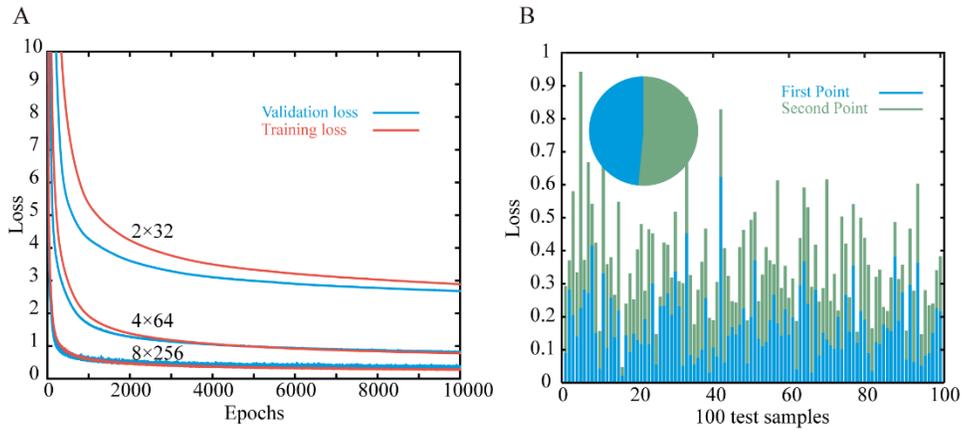

**Figure S5.** A) Training and validation losses during the training process of the 2-point model with structures of 2×32, 4×64, and 8×256. B) The test losses of 100 randomly generated 3-point samples.

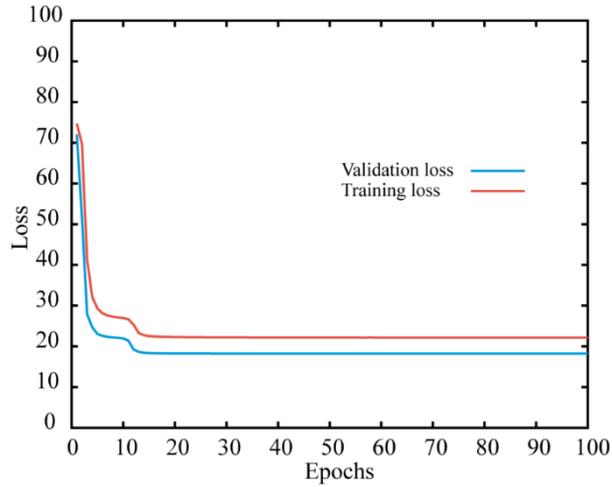

**Figure S6.** A wrong training dataset leads to failed training. In the 2-point model, if ($x_1$, $y_1$, $x_2$, $y_2$) and ($x_2$, $y_2$, $x_1$, $y_1$) are listed in the training dataset simultaneously, the training process will fail as shown above. It is because there two output data are physically the same and have the same input data.

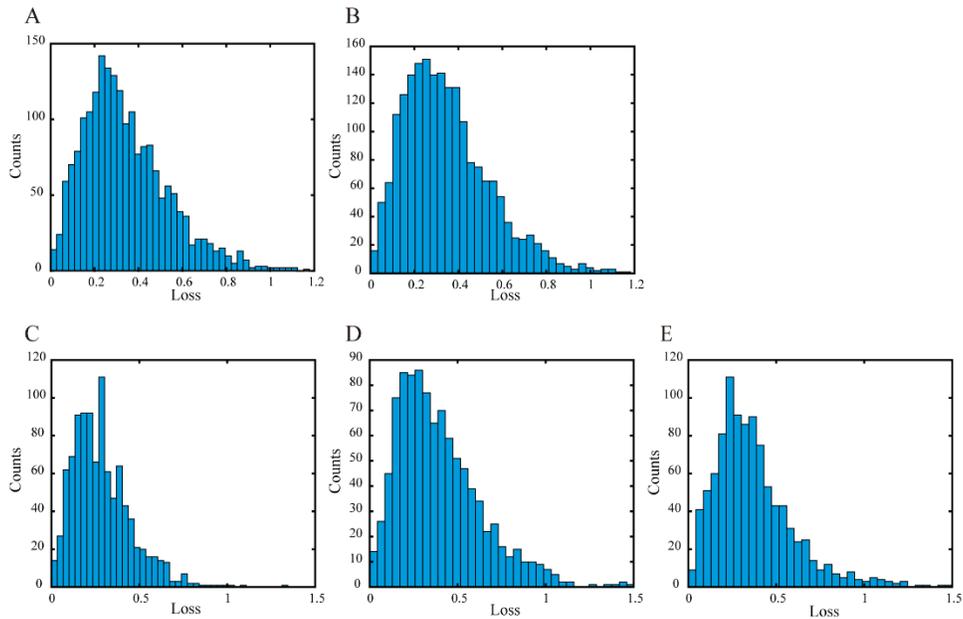

**Figure S7.** Histogram of the test losses of 3000 samples in the test dataset for A-B) 2-point model. C-E) 3-point model.